\documentstyle[12pt,world_sci]{article}
\begin{document}
\begin{flushright}
UM-TH-94-30\\
August 1994  \end{flushright}

\title{{\bf INFRARED EFFECTS AND THE ASYMPTOTICS OF PERTURBATION
THEORY IN WEAK DECAYS OF HEAVY PARTICLES}}
\author{M. BENEKE\thanks{Invited talk presented at the
conference ``QCD'94'', Montpellier, France, July 7 - 13, 1994.
To appear in the proceedings.}\\
{\em Randall Laboratory, University of Michigan, Ann Arbor,\\
Michigan 48109, U.S.A.\\}}

\maketitle
\setlength{\baselineskip}{2.6ex}

\begin{center}
\parbox{13.0cm}
{\begin{center} ABSTRACT \end{center}
{\small \hspace*{0.3cm} I discuss the interplay of infrared
sensitivity in large order
perturbative expansions with the presence of explicit nonperturbative
corrections in the context of heavy quark expansions.
The main focus is on inclusive decays and the status of the kinetic
energy of the heavy quark.
This talk summarizes work done with Braun and Zakharov.
}}
\end{center}

\section{Introduction}
The study of infrared (IR) divergences in perturbation theory (PT)
is crucial to any rigorous approach to hard processes in QCD. The
appearance of an explicitly divergent {\it coefficient} (say,
$\ln\lambda^2$, if a finite gluon mass is used as regulator to leading
order) indicates that the process can not be calculated perturbatively.
However, for a wide class of phenomena, the IR divergence is universal
and can be factorized into a few nonperturbative functions and process
dependent coefficient functions, which are perturbatively calculable.
Defying factorization, in {\it large} orders in PT, the coefficient
function is dominated by IR regions of Feynman integrals, typically
involving a large number of vacuum polarizations in a gluon line. The
corresponding perturbative {\it series} develops an IR renormalon
divergence in large orders, which renders the sum of the series undefined
by terms suppressed by a power of the hard scale. Thus the presence
of IR renormalons implies the existence of ``higher twist'' terms and
requires the introduction of new nonperturbative parameters, such that
the sum of leading and higher twist is unambiguous. [The converse is
not true: Power suppressed terms may exist which are not indicated
by renormalons in previous orders.] In cases where an operator product
expansion (OPE) is available, these parameters are naturally identified
with matrix elements of higher dimension operators, but the argument
is sufficiently general to comprise situations without OPE.

Practically, all present calculations of large order coefficients are
restricted to diagrams with a {\it single} gluon line, dressed by
fermion loops, which is equivalent to integrating the gluon with the
running coupling at the vertex. IR renormalons are then conveniently
discussed in terms of singularities of the Borel transform (BT) $B[\{r_n\}]$
of the series of coefficients $\{r_n\}$ generated in this way. In this
approximation, there is a very transparent relation \cite{BEN94b} between
the singularities of the BT and the low order coefficient
$r_0(\lambda)$, regulated with a finite gluon mass:
\begin{equation}\label{relation}
r_0(\lambda) = \frac{1}{2\pi i}
\!\!\!\int\limits_{-1/2-i\infty}^{-1/2+
i\infty}\!\!\!\!\!\mbox{d} s \,\Gamma(-s)\Gamma(1+s)
\left(\frac{
\lambda^2}{\mu^2} e^C\right)^s\,B[\{r_n\}](s)
\end{equation}
IR renormalon singularities in $B[\{r_n\}]$ are in one-to-one
correspondence with {\it nonanalytic} (in $\lambda^2$) terms
in the small-regulator
expansion of $r_0(\lambda)$. This relation unifies the renormalon
phenomenon with the familiar discussion of explicit IR divergences,
$\ln\lambda^2$. Beyond the restriction to a single gluon line,
a finite gluon mass has to be abandoned as an IR regulator. Within
dimensional regularization one might expect a similar correspondence
of renormalons with poles in different from four dimensions, though
a precise relation analogous to Eq.~(\ref{relation}) has not yet
been established.

In the following, I give a brief summary of results that have been
obtained applying these general ideas to weak decays of heavy hadrons.
In this case the hard scale is provided by the mass of the heavy
quark.

\section{HQET and the pole mass}
Consider the heavy mass expansion of, say, the $B$ meson mass:
\begin{equation}\label{mass}
m_B = m_b^{\rm pole}+\bar{\Lambda}+O\left(1/m_b^{\rm pole}\right)
\end{equation}
The first term is given by the {\it pole} mass of the heavy quark
(HQ), which is therefore the natural expansion parameter for
heavy quark effective theory (HQET). The pole mass is IR finite,
but turns out to be linearly sensitive to IR momenta. Consequently,
the series that relates $m_b^{\rm pole}$ to $m_b^{\overline{\rm MS}}$
(which in principle can be measured to arbitrary accuracy) has an
IR renormalon such that the pole mass is not defined to an accuracy
better than $\Lambda_{\rm QCD}$ within PT \cite{BEN94a,BIG94a}. This
is not unexpected, since $\bar{\Lambda}$ is expected to be of this order.
However, the divergence in the leading term $m_b^{\rm pole}$ implies
that $\bar{\Lambda}$ is not defined by Eq.~(\ref{mass}) by terms of the
same order of magnitude, $\Lambda_{\rm QCD}$,
and only the sum is physical (up to higher orders in $1/m_b^{\rm pole}$).
It follows that HQET does not provide a {\it unique} nonperturbative
definition of the concept of the pole mass.

\section{Exclusive decays}
The parameter $\bar{\Lambda}$ appears (together with new form factors)
in the leading finite mass corrections to the HQ limit of the matrix
elements relevant to exclusive decays. To display the implications
of an ambiguous nature of $\bar{\Lambda}$, the decay $\Lambda_b
\rightarrow \Lambda_c l \bar{\nu}$ is simplest, since finite
mass corrections involve $\bar{\Lambda}$ alone and no new form factors
\cite{GEO90}. The renormalon ambiguity of $\bar{\Lambda}$ is fixed
already by $m_b^{\rm pole}$. Since the physical matrix element must
be unambiguous, a consistency relation emerges: The series of radiative
corrections to the HQ limit must have a renormalon that matches the
ambiguity of $\bar{\Lambda}$. This renormalon arises, because the leading
term in the effective Lagrangian reproduces correctly only the leading
IR contribution, $\ln \lambda^2$, of the full theory matrix element.
Thus, the matching coefficient is IR finite, but contains
$\sqrt{\lambda^2}$. The coefficients have been calculated \cite{NEU94}
and satisfy the consistency relations.

Thus, when $\bar{\Lambda}$ is eliminated in the relation of physical
quantities, no ambiguities remain. On the other hand, if one
attempts to {\it calculate} $\bar{\Lambda}$, e.g. from QCD sum rules
\cite{BAG92}, the ambiguity in the definition Eq.~(\ref{mass}) can not
be avoided and is indeed consistently reflected in the sum rules
\cite{BEN94a}. However, both sum rules and phenomenology point towards
a large value, $\bar{\Lambda}\approx 500\,$ MeV, and one may argue that
this value is larger than the renormalon ambiguity. This is not unreasonable,
because $\bar{\Lambda}$ contains the spectator contribution to the
meson mass in the first place, which is not related to renormalons at all.
{}From this point of view the ambiguity inferred from renormalons can
serve as an intrinsic ``error bar'' on $\bar{\Lambda}$. Whether this picture
is stable numerically, can only be decided by comparison with
phenomenology.

\section{Inclusive decays}
Significant progress has been made over the past years, applying
heavy quark expansions to inclusive heavy flavour decays, e.g., the
semileptonic decay $B\rightarrow X_q l \bar{\nu}$. Within the OPE
and HQET, one finds that $\Lambda_{QCD}/m_b$ corrections are absent
\cite{CHA90,BIG92} and second order corrections can be parameterized
by the kinetic and chromomagnetic energy of the heavy quark inside
the meson, $\mu_K$ and $\mu_G$. The leading term in the decay width
coincides with that of a free quark to all orders in PT and it appears
natural to use the pole mass in
\begin{equation}\label{width}
\Gamma_B  = \frac{G_F^2 \left(m_b^{\rm pole}\right)^5}
{192\pi^3} (1 + {\rm radiative\,\,corr.} + {\rm nonpert.\,\,corr.})\,.
\end{equation}
However, from Sect.~2 above, one concludes that the pole mass has
a large distance ambiguity of order $\Lambda_{\rm QCD}$ in apparent
conflict with the absence of a nonperturbative parameter that could
absorb a $\Lambda_{\rm QCD}/m_b$ correction in Eq.~(\ref{width}).
Thus one might ask whether the short distance expansion (OPE) provides
the correct {\it normalization} of $\Gamma_B$, given a situation,
where a coloured particle (the $b$-quark) lives long in the initial
state. To clarify this question, one has to identify the leading
renormalons (or IR sensitive contributions) in the large order radiative
corrections to the tree decay width $\Gamma_0$ in Eq.~(\ref{width}).
Using Eq.~(\ref{relation}), one may take a finite gluon mass to tag
renormalons and finds \cite{BEN94b}
\begin{equation}\label{spectrum}
\frac{1}{\Gamma_0}\frac{\mbox{d}\Gamma}{\mbox{d}x} =
\Theta(x)\Theta(1-x)\bigg[6 x^2-4 x^3+\frac{\alpha}{3\pi}
\bigg\{F(x) + \frac{\lambda}{m_b} \left(24 x^2-8 x^3\right)
\bigg\}\bigg] + \frac{4\alpha}{3\pi}\frac{\lambda}{m_b}
\,\delta(1-x)
\end{equation}
for the lepton spectrum, where $x=(2 E_l)/m_b^{\rm pole}$, $E_l$ the
lepton energy and $F(x)$ is the one-loop radiative correction for
$\lambda=0$. The $\delta$-function appears, since close to the endpoint
the gluon mass is no longer small compared to the invariant mass
of the hadronic final state. One may now use
$m_b^{\rm pole}=m_b^{\overline{\rm MS}}-(2\lambda\alpha)/3$ to eliminate
the pole mass in favour of a mass parameter that is not linearly
sensitive to large distances (such as $m_b^{\overline{\rm MS}}$) in
Eq.~(\ref{spectrum}). Then all linear in $\lambda$ terms disappear from
the spectrum and consequently the total width, implying cancellation of
the renormalon in the pole mass that could have indicated a
$\Lambda_{\rm QCD}/m_b$ correction  with a renormalon in the radiative
corrections to the free quark decay.

Extending Eq.~(\ref{spectrum}), the subsequent nonanalytic terms
$\lambda^2\ln\lambda^2$ have also been found to vanish \cite{BEN94b}.
In view of Eq.~(\ref{relation}), it follows that the summation of
large order radiative corrections to the free quark decay is free from
ambiguities up to third order in $1/m_b$ or higher. In contrast
to $\bar{\Lambda}$ the kinetic and chromomagnetic energy that appear
in second order are free from renormalon ambiguities. We may argue
that this is true beyond the approximation to which explicit calculations
have been performed. The chromomagnetic energy is related to the
mass splitting of vector and pseudoscalar mesons and trivially free
from ambiguities. For the kinetic energy $\mu_K$ this becomes transparent,
if one visualizes the heavy mass expansion of the width as a two step
process \cite{BIG92}. First one expands the product of two weak Lagrangians
at short distances. The short distance mass is the natural mass parameter
and the kinetic energy operator does not appear. Second one uses HQET
(with the pole mass) to expand the matrix elements. $\mu_K$
arises from the expansion of $\langle B|\bar{b} b|B\rangle/(2 m_B)$,
whose leading term is fixed to unity by current conservation. Thus,
neither step can introduce a series of radiative corrections with a
divergence corresponding to an ambiguity in $\mu_K$. Another way
to see this is to observe that the kinetic energy arises from boosting
the free quark decay to an average frame of the heavy quark inside the
meson. The kinetic energy is therefore protected by Lorentz symmetry
or, in the language of HQET, reparameterization symmetry \cite{LUK92}.

The unambiguous nature of $\mu_K$ is important in two respects: First,
it is a prerequisite to uphold inequalities such as $\mu_K^2 >
\mu_G^2$, that have been derived in various ways \cite{VOL94} in the
presence of renormalization. Second, the calculation of subleading
nonperturbative parameters in HQET on the lattice has faced serious
obstacles in the form of power divergences \cite{MAI92}. While these
in general are closely related to the emergence of renormalons in
the continuum \cite{BEN94a}, for the particular case of $\mu_K$,
mixing with lower dimension operators appears to be a genuine lattice
problem: Reparameterization symmetry which
protects $\mu_K$ in the continuum, is broken on the lattice.

Work supported by the Alexander-von-Humboldt foundation.

\vspace*{0.1cm}
\bibliographystyle{unsrt}

\end{document}